\documentclass{aa} 
\usepackage{graphicx}
\usepackage{natbib,colortbl,graphicx}
\bibpunct{(}{)}{;}{a}{}{,}
\definecolor{lightgray}{gray}{0.8}
\begin{document}
   \title{Habitable planets around the star Gl~581?}

   \author{F. Selsis
          \inst{1,2}
          \and
          J. F. Kasting
          \inst{3}
          \and
          B. Levrard
          \inst{4,1}
          J. Paillet
          \inst{5}
          \and
          I. Ribas
          \inst{6}
          \and
          X. Delfosse
          \inst{7}
           }
   \offprints{F. Selsis}
\institute{
CRAL: Centre de Recherche Astrophysique de Lyon (CNRS; Universit\'e de Lyon; Ecole Normale Sup\'erieure de Lyon), 46 all\'ee d'Italie, F-69007, Lyon, France,  \\
\email{franck.selsis@ens-lyon.fr}              
\and
LAB: Laboratoire d'Astrophysique de Bordeaux (CNRS; Universit\'e Bordeaux I), BP 89, F-33270 Floirac, France
\and
Dept. of Geosciences, The Pennsylvania State University, University Park, Pennsylvania 16802, USA, \\
\email{kasting@geosc.psu.edu}   
\and
IMCCE: Institut de M\'ecanique C\'eleste et de Calcul des Eph\'em\'erides (CNRS ; Universit\'e Pierre et Marie Curie - Paris VI), 77 Avenue Denfert-Rochereau, F-75014, Paris, France, 
\email{Benjamin.Levrard@imcce.fr}
\and
ESA/ESTEC SCI-SA, Keplerlaan 1, PO BOX 299, 2200AG Noordwijk, The Netherlands,       
\email{jpaillet@rssd.esa.int}  
\and
Institut de Ci\`encies de l'Espai (CSIC-IEEC), Campus UAB, 08193 Bellaterra, Spain, 
\email{iribas@ieec.uab.es}
\and
LAOG: Laboratoire d'AstrOphysique de Grenoble, (CNRS; Universit\'e J. Fourier - Grenoble I), BP 53X, 38041 Grenoble Cedex, France, 
\email{delfosse@obs.ujf-grenoble.fr}
}

   \date{Received June 15, 2007; accepted October 26, 2007}

 
  \abstract
      {Thanks to remarkable progress, radial velocity surveys are now able
to detect terrestrial planets at habitable distance from low-mass stars.
Recently, two planets with minimum masses below 10~$M_{\oplus}$have been
reported in a triple system around the M-type star Gliese 581. These
planets are found at orbital distances comparable to the location of the
boundaries of the habitable zone of their star. }
     {In this study, we assess the habitability of planets Gl~581c and Gl
581d (assuming that their actual masses are close to their minimum masses)
by estimating the locations of the habitable-zone boundaries of the star
and discussing the uncertainties affecting their determination. An
additional purpose of this paper is to provide simplified formulae for
estimating the edges of the habitable zone. These may be used to evaluate
the astrobiological potential of terrestrial exoplanets that will
hopefully be discovered in the near future.}
     {Using results from radiative-convective atmospheric models and
constraints from the evolution of Venus and Mars, we derive theoretical
and empirical habitable distances for stars of F, G, K, and M spectral
types. }
     {Planets Gl~581c and Gl~581d are near to, but outside, what can be
considered as the conservative habitable zone. Planet `c' receives 30\%
more energy from its star than Venus from the Sun, with an increased
radiative forcing caused by the spectral energy distribution of Gl~581.
This planet is thus unlikely to host liquid water, although its
habitability cannot be positively ruled out by theoretical models due to
uncertainties affecting cloud properties and cloud cover. Highly
reflective clouds covering at least 75\% of the day side of the planet
could indeed prevent the water reservoir from being entirely vaporized.
Irradiation conditions of planet `d' are comparable to those of early
Mars, which is known to have hosted surface liquid water.  Thanks to the
greenhouse effect of CO$_2$-ice clouds, also invoked to explain the
early Martian climate, planet `d' might be a better candidate for the
first exoplanet known to be potentially habitable. A mixture of several
greenhouse gases could also maintain habitable conditions on this planet,
although the geochemical processes that could stabilize such a
\textit{super-greenhouse} atmosphere are still unknown. }
   {}


   \maketitle
 \section{Introduction}

The M-type star Gl~581 hosts at least 3 planets, which were detected using
radial velocity measurements by Bonfils et al.
\citeyearpar{2005A&A...443L..15B} (planet 'b') and Udry et al.
\citeyearpar{2007A&A...469L..43U} (planets `c' and `d'). The properties of
this star and its planets are given in Table~\ref{tab:gl581}. Before this
discovery, only two exoplanets were known to have a minimum mass below
10~$M_{\oplus}$, which is usually considered as a boundary between
terrestrial and giant planets, the latter having a significant fraction of
their mass in an H$_2$-He envelope. The first one was GJ~876d, a very hot
planet ($P\leq 2$~days) with a minimum mass of 5.9~M$_{\oplus}$
\citep{2005ApJ...634..625R}. The other one is OGLE-05-390L b, found to be
a $\sim$5.5~M$_{\oplus}$ cold planet at 2.1~AU from its low-mass
parent star thanks to a microlensing event
\citep{2006Natur.439..437B,2006ApJ...651..535E}. Neither of these two
planets is considered as habitable, even with very loose habitability
criteria. In the case of Gl~581, and as already mentioned by Udry et al.
(2007), the locations of planet `c' and `d' must be fairly close to the
inner and outer edges, respectively, of the habitable zone (HZ). In
this paper, we investigate the atmospheric properties that would be
required to make the habitability of these planets possible.
  
Because of its equilibrium temperature of $\sim$300~K when calculated with
an albedo of 0.5, it has been claimed that the second planet of this system, Gl~581c, is potentially habitable (Udry et al. 2007), with climatic
conditions possibly similar to those prevailing on Earth. After a brief
discussion about the relationship between the equilibrium temperature and
habitability, we summarize in this paper what are usually considered as
the boundaries of the circumstellar HZ and the uncertainties on their
precise location.  In Sect.~\ref{sec:HZstars} we provide
parameterizations to determine such limits as a function of the stellar
luminosity and effective temperature.  These can be used to evaluate the
potential habitability of the terrestrial exoplanets that should soon be
discovered. We then discuss the specific case of the system around Gl~581.


 \begin{table}[!t]
\caption{Properties of the star Gl~581 and its 3 detected planets, from
Udry et al. (2007).}
\begin{center}
\begin{tabular}{lllll}
  \hline
Star & $T_{\rm eff}$ (K) &$M$/M$_{\odot}$ & $R$/R$_{\odot}$ &  
$L$/L$_{\odot}$ \\
  \hline
  Gl~581 & 3200 &0.31 & 0.38 & 0.0135 \\
  \hline
  &&& \\
 &&& \\
  \hline
 Planets  & $a$~(AU) & $M_{\rm min}$/M$_{\oplus}$ & $R_{\rm min}$/R$_{\oplus}$ & stellar flux\\
  &   &$^{*}$ & $^{**}$  & $S/S_{0}$$^{***}$\\
  \hline
  b & 0.041 & 15.6 & 2.2-2.6 & 8.1 \\
\rowcolor{lightgray}
 c & 0.073 & 5.06 & 1.6-2.0 & 2.55 \\
 \rowcolor{lightgray}
 d & 0.253 & 8.3 & 1.8-2.2 & 0.21 \\
  \hline
\end{tabular}
\end{center}
The potential habitability of planets `c' and `d',
highlighted in grey, is discussed in this paper. \\
$^{*}$ $M_{\rm min}=M \sin i$, where $i$ is the orbital inclination.\\
$^{**}$ Radius for a rocky and ocean planet, respectively
\citep{2007Sotin,2007ApJ...665.1413V}.\\
$^{***}$ $S_0$ is the solar flux at 1 AU: 1360 W m$^{-2}$.\\
\label{tab:gl581}
\end{table}

\section{Habitable planets and the habitable zone}

The HZ is the circumstellar region inside which a terrestrial planet can
hold permanent liquid water on its surface. A terrestrial planet that is
found beyond the HZ of its star could still harbor life in its subsurface;
but being unable to use starlight as a source of energy, such endolithic
biosphere would not be likely to modify its planetary environment in an
observable way \citep{2005IJAsB...4....9R}. In the Solar System,
{\em in situ} searches for biological activity in the subsurface of,
for instance, Mars or Jupiter's satellite Europa could in principle be
carried out. But with exoplanets presumably out of reach for {\em in
situ} exploration, signs of life will have to be searched via signatures
of photosynthetic processes in the spectra of planets found in the HZ of
their stars. This is the purpose of future space observatories such as
Darwin \citep{darwin2000,2005AdSpR..36.1114K}, TPF-C \citep{TPFC2006} and
TPF-I \citep{1999tpf}.  For exoplanets, ``habitable'' thus
implies {\it surface habitability}.

A planet found in the HZ is not necessary habitable. The maintenance
of habitable conditions on a planet requires various geophysical and
geochemical conditions. Only some of them, those that have a direct
influence on the atmospheric properties, are addressed in the present
paper \citep[see for instance][for a comprehensive view of
habitability]{2007AsBio...7...85S,2007SSRv..tmp..127Z,2003ARA&A..41..429K,1999eehw.book.....L,2007prpl.conf..929G}.
Many factors may prevent (surface) habitability. To give several examples: 
the planet may lack water, the rate of large impacts may be too high, the
minimum set of ingredients necessary for the emergence of life (so far
unknown) may have not been there, gravity may be too weak (as on Mars) to
retain a dense atmosphere against escape processes and to keep an active
geology replenishing the atmosphere of CO$_2$, or the planet could have
accreted a massive H$_2$-He envelope that would prevent water from being
liquid by keeping the surface pressure too high. To avoid the two last
scenarios, the planetary mass should be in the approximate range of
0.5--10~$M_\oplus$, although this is more of an educated guess than a
reliable quantitative estimate.

\begin{figure}[t]
\begin{center}
\includegraphics[width=\linewidth]{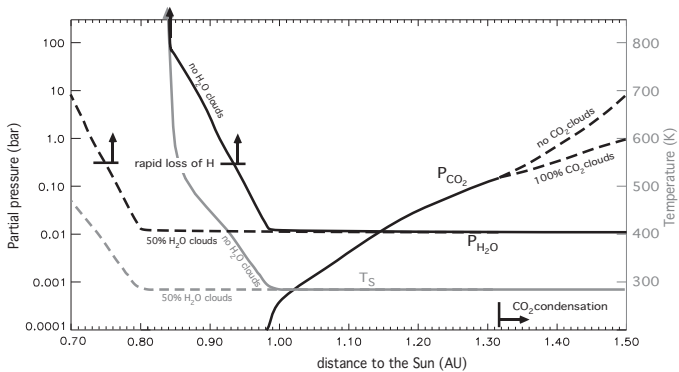}
\caption{The CO$_2$ and H$_2$O pressure and the mean surface temperature
of a habitable planet as a function of the orbital distance. The
diagram gives $P_{{\rm CO}_2}$ and P$_{{\rm H}_2{\rm O}}$ (left y-axis)
and $T_S$ (grey, right y-axis) of an Earth-like planet across the inner
part of the HZ around the present Sun. At orbital distances
larger than $\sim0.98$~AU, $T_S$ is assumed to be fixed at 288~K (its
current value on Earth) by the carbonate-silicate cycle, which is
obviously an idealized picture. Beyond 1.3~AU, CO$_2$ condenses in the
atmosphere and the required level of CO$_2$ depends on the coverage by
CO$_2$-ice clouds. The pressure of H$_2$O and $T_{\rm s}$ in the inner HZ
is calculated for a cloud-free atmosphere and 50\% cloud cover and
assuming a reservoir of water that contains (as on Earth) more than the
equivalent of 220~bars of H$_2$O. }
\label{fig:P_T}
\end{center}
\end{figure}

 Being at the right distance from its star is thus only one of the
necessary conditions required for a planet to be habitable. In the current
absence of observational constraints, we choose to assess the habitable
potential of the planets with as few hypotheses as possible on their
physical and chemical nature. We therefore assume that the planet
satisfies only two conditions. Although these two conditions are very
simple, they may derive from complex geophysical properties. Future
observations will hopefully tell us whether such properties are frequent
or rare on terrestrial exoplanets. These conditions are:
\begin{itemize}
\item[$i)$] The amount of superficial water must be large enough so that the
surface can host liquid water for any temperature between the temperature
at the triple point of water, 273~K, and the critical temperature of
water, $T_{\rm{c}}$=647~K. This condition implies that the water reservoir
produces a surface pressure higher than 220~bars when fully vaporized.
With an Earth gravity, this corresponds to a 2.2~km layer of water,
slightly lower than the mean depth of Earth oceans of 2.7~km. For a
gravity twice that of Earth, this pressure corresponds to half
this depth. Planets with less water may still be habitable, but their HZs
may be somewhat narrower than we calculate here because liquid water would
disappear at a lower surface temperature.
\item[$ii)$]  Atmospheric CO$_2$ must accumulate in a planet's atmosphere
whenever the mean surface temperature falls below 273~K, the freezing
point of water. This is a consequence of the carbonate-silicate cycle,
which stabilizes the long-term surface temperature and the amount of
CO$_2$ in the atmosphere of the Earth \citep{1981jgrW}. Such an assumption
implies that the planet is geologically active and continuously outgassing
CO$_2$. It also implies that carbonates form in the presence of surface
liquid water, which may require continental weathering. With no
atmospheric CO$_2$, or with a fixed CO$_2$ level as in Hart
\citeyearpar{1979Icar...37..351H}, the HZ could be $\sim$10 times narrower
than is currently assumed. In the absence of CO$_2$ (or a greenhouse gas
other than H$_2$O), the present Earth would be frozen.
\end{itemize}
The atmosphere of a habitable planet meeting these conditions should
behave as illustrated in Fig~\ref{fig:P_T}.

\subsection{The equilibrium temperature}

The equilibrium temperature of a planet is given by 
\begin{equation}
T_{\rm eq} = \left ( \frac{S(1-A)}{f \sigma} \right ) ^\frac{1}{4} 
\end{equation}
where $A$ is the {\it Bond} albedo (which is the fraction of power
{\em at all wavelengths} scattered back out into space -- Earth's value
is 0.29), $S$ is the stellar energy flux, $\sigma$ is the Stefan-Boltzmann
constant, and $f$ is a redistribution factor. If all the incident energy
is uniformly distributed on the planetary sphere, then $f=4$. If the
energy is uniformly distributed over the starlit hemisphere alone, then
$f=2$. And if there is no redistribution, the {\em local} equilibrium
temperature can be calculated with $f=1/\cos \theta$ where $\theta$ is the
zenith angle. The latter case, for instance, yields good results for the
surface temperature on the sunlit hemispheres of airless bodies with known
albedo such as the Moon and Mercury.

It is important to discuss the meaning of $T_{\rm eq}$ and the manner
in which it can be used to assess habitability. The planet Gl~581c has
been widely presented as potentially habitable because one finds $T_{\rm
eq} \sim 320$~K when calculated using the albedo of the Earth. This
conclusion is however too simplistic for the following two reasons:
\begin{itemize}
\item[$i)$] For a planet with a dense atmosphere (an inherent
property of a habitable planet), $T_{\rm eq}$ does not indicate any
 physical temperatures at the surface or in the atmosphere.
With albedos of 0.75, 0.29, and 0.22, respectively, and assuming $f=4$,
Venus, Earth, and Mars have equilibrium temperatures of 231~K, 255~K, and
213~K, while their mean surface temperatures are 737~K, 288~K and 218~K.
The two quantities only match, approximately, in the case of Mars, whose
tenuous atmosphere produces a greenhouse warming of only $\sim 5$~K.
\item[$ii)$] It can be demonstrated that a necessary (but not sufficient)
condition for habitability is that $T_{\rm eq}$ must be lower than about
270~K. If the surface temperature remains below the critical temperature
of water ($T_{\rm c}=647$~K), the thermal emission of a habitable planet
cannot exceed the runaway greenhouse threshold, $\sim300$~W m$^{-2}$ (see
Sect.~\ref{sec:RG}), equivalent to the irradiance of a black-body at
270~K. Therefore, if a planet has an atmosphere and an equilibrium
temperature above 270~K, two situations may arise. First, $T_{\rm s}$ may
remain below $T_{\rm c}$, but there would be no liquid water at the
surface and no or negligible amounts of water vapor in the atmosphere. In
a second possible situation, the atmosphere contains considerable amounts
of water vapor, but the surface temperature exceeds $1400$~K (see
Sect.~\ref{sec:RG}). This would allow the planet to balance the absorbed
stellar energy by radiating at visible and radio wavelengths through an
atmosphere that is optically thick in the infrared (IR). Both cases would
render the planet uninhabitable.
\end{itemize}
For planet Gl~581c to be habitable (i.e., $T_{\rm eq}<270$~K), its albedo
would have to be higher than 0.65. Since planet Venus has an albedo of
0.75, this situation may not appear unrealistic. However, as we will see
in the next sections, the Bond albedo is not a quantity given by the
planetary characteristics alone, but the spectral energy distribution of
the star also needs to be taken into account.

\subsection{The inner edge of the HZ}\label{sec:inner}

Let us consider a planet with a large water reservoir covering its entire
surface, but no other greenhouse volatiles. As a first step, we assume that
its host star is a Sun-like star and that the planet has the same gravity
as the Earth. For a given orbital distance, a fraction of the water
reservoir is in the form of vapor. The surface temperature $T_{\rm s}$
imposes the surface vapor pressure $P_{\rm w}$. If $P_{\rm w}$ is high
enough, the water vapor, in turn, affects $T_{\rm s}$ by blocking the
outgoing IR radiation, by reducing the atmospheric lapse rate and by
modifying the planetary albedo. To account for this coupling, the
atmospheric structure has to be computed self-consistently for a given
irradiation. This was done previously by using a 1D radiative-convective
model (Kasting 1988). All the orbital distances and stellar fluxes in the
following subsections are given relative both to the present Sun and to
the present solar flux at Earth orbit ($S_0=1360$~W m$^{-2}$). We will see
further how these values can be scaled to other stellar luminosities and
effective temperatures. Values given in Sect.~\ref{sec:RG},
\ref{sec:waterloss}, and \ref{sec:thermophilic} were obtained by Kasting
(1988) with a cloud-free radiative convective scheme.
Section~\ref{subsec:clouds} discusses the likely effects of clouds.

\subsubsection{The runaway greenhouse limit}\label{sec:RG}

For orbital distances smaller than 1 AU (and for the present solar
luminosity), $T_{\rm s}$ is extremely sensitive to the orbital distance,
increasing from less than 273~K at 1 AU (in the absence of CO$_2$) to
about 373~K ($P_{\rm w}=1$~bar) at 0.95 AU (see Fig.~\ref{fig:P_T}). This
sharp increase in $T_{\rm s}$ is mainly caused by the increase in the IR
opacity and the decrease in the albedo caused by absorption of solar
near-infrared (NIR) radiation by water vapor. For even smaller orbital
distances, because of the relation between temperature, vapor pressure, and
IR opacity, the outgoing IR flux becomes nearly independent of the surface
temperature and tends asymptotically towards its upper limit of about
300~W m$^{-2}$, known as the runaway greenhouse threshold
\citep{1988JAtS...45.3081A,1988Icar...74..472K,2002JAtS...59.3223I}. At
this point, an increase in the irradiation (or a decrease of the orbital
distance) does not result in an increase in the outgoing IR flux, but
leads instead to a strong increase in $T_{\rm s}$ and $P_{\rm w}$. In
turn, this produces a slight increase in the albedo and thus in the
reflected visible/NIR radiation. The increase in the albedo for $T_{\rm
s}$ above 373~K and $P_{\rm w}$ above 1~bar is the consequence of the
strong Rayleigh back-scattering occurring in the visible (a spectral
domain where water vapor does not absorb significantly) and to the
saturation of the water bands absorbing the stellar NIR radiation. This
increase in the atmospheric albedo, up to about 0.35 (in the absence of
clouds), protects the water reservoir from complete vaporization for
orbital distances down to 0.84~AU (1.4 $S_0$). At 0.84~AU, $T_{\rm s}$
reaches $T_{\rm c}=647$~K, and the water reservoir becomes a supercritical
fluid envelope. A more limited water reservoir could, of course, be fully
vaporized at lower irradiation. When this theoretical limit for the
irradiation is crossed, there is a dramatic increase in $T_{\rm s}$ from
$>647$~K to $>1400$~K, a temperature that potentially melts silicates on
the surface. This behavior is a consequence of the runaway greenhouse
threshold that limits the mid-IR cooling of the planet. At $T_{\rm
s}>1400$~K, the planet can radiate the absorbed stellar energy through the
atmosphere at the visible and radio wavelengths, at which the water vapor
opacity is negligible.

\subsubsection{The water loss limit}\label{sec:waterloss}

In a cloud-free radiative-convective scheme, water vapor would become a
major atmospheric constituent in an Earth analog placed at 0.95~AU from
the present Sun. The loss of hydrogen to space would no longer be limited
by the diffusion of water vapor from the troposphere to the stratosphere,
but by the stellar EUV energy deposited in the upper atmosphere, and would
be enhanced by $\sim$4 orders of magnitude. The hydrogen contained in the
whole terrestrial ocean would thus be lost in less than 1~Gyr, which would
terminate Earth's habitability. For these two reasons, 0.95~AU could be
seen as the inner limit of the present solar HZ.

The water loss limit is difficult to extrapolate to other stars and
terrestrial exoplanets.  Such a limit for present Earth corresponds to a
surface temperature of $\sim$340~K.  This is because the background
atmospheric pressure is 1~bar (mainly N$_2$ and O$_2$), and the vapor
pressure of H$_2$O at 340~K is 0.2 bar, making H$_2$O a 20\% constituent
of the atmosphere. At this H$_2$O mixing ratio, the loss of H to space
becomes energy-limited.  If the background surface pressure is higher,
then the water-loss limit will be reached at a higher temperature. For
instance, with a background atmospheric pressure of 5 bars, the water loss
limit would be reached at about 373~K. In addition, the water content of
terrestrial planets is thought to be highly variable
\citep{2007AsBio...7...66R} and could be as high as 50\% in mass for
migrating planets initially formed beyond the snow line
\citep{2003ApJ...596L.105K,2004ocean_planets,selsis_OPs_2007}. For
such water-rich planets, only a fraction of the water can be lost within
the lifetime of the planet, and atmospheric escape is not a threat to
habitability.

Planets in the Gl~581 system are good {\em ocean planet} candidates
because the architecture of the system and the high mass of the planets
are likely to be inherited from type-I migration. These planets can thus
be initially composed of a large fraction of water ice. In addition, the
lifetime of the water reservoir also depends on the stellar emission in
the UV, which photodissociates H$_2$O, and in the XUV (0.1--100~nm), which
produces thermospheric heating. The star Gl~581 appears quite inactive and
should not have strong UV or XUV fluxes presently. Thus, gravitational
escape is not likely to play a major role in the system today. However,
the high-energy emissions of Gl~581 may have been orders of magnitude
stronger in the past. The evolution of these
emissions has not been accurately established for M-type stars yet, but
they are likely to have had an impact on the planet's atmosphere for a
more or less extended period of time. This point is addressed in more
detail in Sect. \ref{sec:erosion}

As already mentioned, the HZ comprises the orbital regions where
terrestrial planets can be probed in search of biosignatures. In this
context, water loss is an important issue, because the leftover oxygen could
possibly produce a dense O$_2$-rich atmosphere. This may be impossible to
distinguish from an atmosphere sustained by photosynthesis, unless
additional biomarker gases were also detected. Therefore, planets found
closer to their parent star than the water-loss limit represent
questionable astrobiological targets.

\subsubsection{The thermophilic limit}\label{sec:thermophilic}

Terrestrial organisms have the extraordinary ability to adapt themselves
to extreme conditions, including hot environments. For hyperthermophilic
prokaryotes, the optimum temperature for growth is above 353~K.
{\em Pyrolobus fumarii}, an iron-breathing bacterium, has been found to
tolerate temperature as high as 394~K, setting the record for the highest
temperature known to be compatible with life. This temperature limit is
lower ($\sim$333~K) for eukaryotes and photosynthetic prokaryotes. The
orbital distance at which $T_{\rm s}$ cannot be lower than 394~K
could thus be seen as an empirical, anthropocentric, inner edge of the HZ.
Interestingly, this highest temperature tolerated by life (as we know it)
is close to the mean surface temperature at which the loss of water
becomes considerable, so that this limit and the water-loss limit are
located at approximately the same orbital distance of $\sim$0.94 AU for a
cloud-free planet.

\subsubsection{The Venus criterion}

Radar maps of the surface of Venus suggest that liquid water has not been
present there for at least the last 1~Gyr \citep{1991Sci...252..252S}. The
Sun was $\sim$8\% dimmer at that time, according to standard solar evolution
models (e.g., Baraffe et al. \citeyear{1998baraffe}). Thus, the solar flux
at Venus' orbit was then equal to what it is today at a distance of 0.72~AU$\times$
(1/0.92)$^{0.5}\sim$0.75~AU. This provides an empirical indication
of the location of the inner edge of the HZ. We do not know if Venus lost
its water content after experiencing a runaway greenhouse or because its
water reservoir was much smaller than Earth's. The D/H ratio measured in
Venusian water vapor traces is $120\pm40$ times higher than on Earth
\citep{1991Sci...251..547D}. Thus, if the initial D/H ratio of Venusian
water was the same as on Earth, there must have been at least $\sim$120
times more water on early Venus than today. This corresponds to an initial
inventory of about 20~m of precipitable water. Since deuterium is also
lost to space, albeit at a slower rate than H, and volcanoes release
poorly deuterated hydrogen into Venus' atmosphere, this lower limit can
significantly underestimate the initial water content. This reveals that
the cold trap limiting the flux of water vapor from the troposphere to the
stratosphere did not work on Venus as it worked on Earth. In turn, this
implies that 1~Gyr ago or longer the mean surface temperature was high
enough to trigger massive water loss ($\sim$340~K or higher depending of
the background atmospheric pressure).

Is it possible to explain the D/H ratio of Venus atmosphere if we
assume that the current CO$_2$-rich, hot, and mainly dry conditions have
always been prevailing at the surface of Venus? The answer is yes, if the
deuterium enrichment has been generated by the loss of the water delivered
sporadically by impacts. The minimum amount of water lost
($\sim10^{16}$~m$^3$ in volume, $10^{19}$~kg in mass) corresponds to the
contents of 25\,000 Halley-sized comets. Venus should not have experienced
such an accumulated delivery since the late heavy bombardment (LHB), which
occurred 3.95--3.85~Gyr ago. Through numerical simulations, Gomes et al.
\citeyearpar{2005Natur.435..466G} found that $10^{19}$~kg of cometary
material ($\sim5\times10^{18}$~kg of water) impacted on the Moon during
the LHB.  When scaling this total impact volume to Venus (by assuming a
factor 1--10 increase relative to the Moon) and taking the
cometary D/H ratio (twice as large as the terrestrial ocean value) into account, we can
see that enough water could have been brought to Venus during the LHB.
However, if this water was lost soon after the LHB, deuterium has then
been escaping during the last 3.85~Gyr and only a small fraction should
remain today. This implies a much higher water delivery that is unlikely
to be consistent with post-accretion impacts. In summary, although
impact-delivered water is unlikely to explain the present D/H enrichment
of Venus, we must keep in mind that the Venus criterion assumes that the
lost water was condensed on the surface at some point during Venus'
history, which has not been proved.

\subsubsection{The effect of clouds on the inner boundary of the
HZ}\label{subsec:clouds}

On a habitable planet close to the inner boundary of the HZ, the IR
opacity of the atmosphere is fully dominated by water vapor, and clouds do
not contribute to warming the surface, as some types of clouds (high
cirrus clouds) can do on Earth. But, on the other hand, clouds can
significantly increase the planetary albedo and thereby reduce the
greenhouse warming. In particular, thick clouds forming at high altitude,
above the level where the incident radiation is backscattered or absorbed,
can result in a very high albedo and can thus move the habitability limits
closer to the star. The 1-D radiative convective models that have been
used to estimate the climatic response to an increase of stellar flux
\citep{1988Icar...74..472K,1991Icar...94....1K} can only bracket the
quantitative effects of clouds on the planetary radiation budget. Clouds
are by nature a 3-D phenomenon that is closely related to atmospheric
circulation. By adding a cloud layer in a 1-D model, it is however
possible to investigate the effect of a 100\% cloud cover. We can also
estimate the value of $T_{\rm s}$ with a 50\% cloud coverage (or any other
percentage) by assuming that clouds do not affect the IR outgoing
radiation. This allows us to calculate the total albedo by combining
cloud-free and cloudy models and to find the orbital distance at which the
absorbed energy matches the IR outgoing radiation. This approximation is
acceptable for large enough water vapor columns, which are found for
$T_{\rm s}$ roughly above 373~K, and thus for calculations near the inner
edge. The $T_{\rm s}$ and $P_{\rm w}$ curves for the 50\% cloud coverage
case in Fig.~\ref{fig:P_T} are thus reliable only in the innermost part of
the HZ (for $T_{\rm s}>373$~K).

A cloud layer located between the 0.1 and 1~bar levels (just above the
surface in the case of present Earth, but at an altitude of $\sim$150~km
for a hotter planet with more than 100 bar of H$_2$O) produces the maximum
increase in the albedo. For a planet orbiting our present Sun, these
clouds can produce an albedo as high as 0.8 and 0.6, respectively, when
covering 100\% and 50\% of the day side. This would move the runaway
greenhouse limit to 0.46~AU (100\%) and 0.68~AU (50\%) from the Sun.
Interestingly, the water-loss limit ($T_{\rm s}=373$~K) with 50\%
cloudiness matches the empirical Venus limit (0.72~AU). Such high-altitude
and thick clouds form mainly by condensation in updrafts, and hence are
unlikely to cover a planet's entire hemisphere. Joshi
\citeyearpar{2003AsBio...3..415J} simulated the 3-D climate of a
synchronously rotating planet (which is expected for planets in circular
orbits within the HZ of M stars, see Sect.~\ref{sec:sync}), with various ocean/continent ratios. The
author found that the high-altitude clouds were mainly located on the dark
side of the planet and that the cloud cover on the day side was below
50\%. This simulation was calculated for a thin Earth-like atmosphere
and should not be used to constrain the cloudiness of a thick H$_2$O-rich
atmosphere close to the inner edge of the HZ. It is, however, a good
illustration of the simulations that should be developed in the future to
address the meteorology of exoplanets.

\subsection{The outer edge of the HZ}

When the carbonate-silicate cycle is at work, the level of CO$_2$ should
be stabilized at a value sustaining a mean surface temperature $T_{\rm s}$
somewhere above 273~K. It has to be more than a few degrees above 273~K
because of the runaway ice-albedo feedback that would trigger a global
glacial event if a relatively large fraction of the oceans is frozen. The
precise stabilizing temperature should depend on the planet's land/ocean
distribution, internal heat flow, surface carbon inventory, and other
factors that are difficult to constrain and that may vary from planet to
planet.

Note that models computing the fluxes between the different CO$_2$
reservoirs (atmosphere, ocean, biosphere, crust, mantle) and the
consequent CO$_2$ atmospheric level on an Earth-like planet do exist
\citep{2002TellB..54..325F,2001lowco2}. Unfortunately,
such models typically depend on planetary parameters (bulk composition,
volatile content, radiogenic elements abundances, structure, formation
history, internal energy, heat flux,...) that are impossible
to estimate before a planet has actually been discovered and
deeply characterized.



Considering the limited understanding of the carbon cycle on Earth, and
the expected diversity of exoplanets, we make the very simple
assumption that the CO$_2$ level is potentially maintained at a habitable
level. If it is not, then the planet is obviously not habitable, which
recalls that being inside the habitable zone is a necessary
but not sufficient condition for habitability.

The relationship between the orbital distance and the required level of
CO$_2$ is given in Fig.~\ref{fig:P_T}, where the stabilization temperature
is arbitrarily fixed to 288~K, the current value of $T_{\rm s}$ on Earth.
At 1~AU from the present Sun, a CO$_2$ abundance of $3\times 10^{-4}$ bar
is sufficient to keep $T_{\rm s}$ of the Earth at 288~K. Four billion
years ago, the solar luminosity was about 70\% of its present value, and
1000 times more CO$_2$ should have been required to maintain the same
$T_{\rm s}$. If the Earth was placed at 1.2~AU from the present Sun, the
CO$_2$ level would stabilize at approximately this same high value. The
farthest orbital distance at which the Earth could be placed without
freezing permanently represents the outer boundary of the HZ. This outer
edge has not been accurately determined yet because of the complex effects
of clouds resulting from CO$_2$ condensation.

If we consider a cloud-free CO$_2$ atmosphere (with a water pressure fixed
by $T_{\rm s}$), the outer edge should lie at 1.67~AU for the present Sun,
where a planet would have a CO$_2$ pressure of about 8~bar (Kasting 1991).
For CO$_2$ pressures above 6~bar, the cooling caused by the albedo exceeds
the warming caused by the IR opacity of the CO$_2$ column. However, this
1.67~AU limit does not take into account the effect of CO$_2$ clouds,
which should form a significant cover at orbital distances larger than
1.3~AU.

\subsubsection{The CO$_2$-cloud limit}

The optical properties of CO$_2$-ice clouds differ significantly from
those made of water droplets or H$_2$O-ice particles. Carbon dioxide
clouds are more transparent in the visible range, but they efficiently
scatter thermal radiation around 10~$\mu$m. Forget \& Pierrehumbert
\citeyearpar{1997Sci...278.1273F} showed that the warming effect caused by
the backscattering of the IR surface emission exceeds the cooling effect
caused by the increase of albedo. As a consequence, CO$_2$ condensation
increases the greenhouse warming compared to a purely gaseous CO$_2$
atmosphere. Depending on the fractional cloud cover, the theoretical outer
edge of the HZ should be found between the 1.67~AU cloud-free limit and a
100\% cloud cover at 2.4~AU
\citep{1997Sci...278.1273F,2000Icar..145..546M}.

\subsubsection{The early Mars criterion}

Numerous geological and geochemical features indicate that liquid water
was present on the surface of Mars as early as 4~Gyr ago
\citep{1987Icar...71..203P,2006Sci...312..400B}, when the luminosity of
the Sun was 28\% lower. The solar flux at Mars' orbit was then equal to
what it is today at an orbital distance of 1.5~AU$\times$ (1/0.72)$^{0.5}\sim$1.77 AU.
Whatever the cause of the greenhouse warming on early Mars, this
fact suggests empirically that the outer edge is located beyond this
distance. A likely explanation for the early habitability of Mars is the
climatic effect of CO$_2$ clouds, perhaps combined with additional warming
by reduced greenhouse gases. Most of the geological features associated
with running water can indeed be explained by a 3-D model of the early
Martian climate, including clouds, with a $\sim1$~bar CO$_2$ atmosphere
\citep{2007leas.book..103F}. Therefore, the outer boundary of the solar HZ
must be located somewhere between 1.77 and 2.4~AU.

\subsubsection{A greenhouse-cocktail limit?} \label{sec:super-greenhouse}

If a planet's atmosphere contains other greenhouse gases in addition to
H$_2$O and CO$_2$, surface conditions could remain habitable at larger
orbital distances than the outer limit defined by the warming of CO$_2$
clouds. Reduced gases, such as methane (CH$_4$) or ammonia (NH$_3$),
could for instance increase the greenhouse effect
\citep{1997Sci...276.1217S}. Methane could have been an important
greenhouse gas on early Earth when oxygen was only a trace constituent of
the atmosphere. At that time, the photochemical lifetime of atmospheric
CH$_4$ was longer and a release of this gas at its present rate would have
resulted in an atmospheric level 100--1000 times higher than today. Such
a high level of CH$_4$ produces a significant greenhouse warming by
absorbing the thermal IR in the 7.4~$\mu$m band. The efficiency of CH$_4$
as a greenhouse gas has, however, been notably revised downwards by Haqq-Misra
et al.  \citeyearpar{Haqq_Misra_CH4} compared to earlier estimates by
Pavlov et al.  \citeyearpar{2000JGR...10511981P}. The IR absorption of
CH$_4$ is observable in a global low-resolution mid-IR spectrum, as could
be measured by Darwin or TPF-I \citep{2007ApJ...658..598K}. In fact, a
spectrum of a planet's IR emission can potentially reveal any atmospheric
species working as a major greenhouse gas.

Since CO$_2$ is not the only greenhouse gas able to maintain habitable
conditions at low stellar irradiation, one could use atmosphere models to
find what the optimum composition is to produce a mean surface temperature
above 273~K at the largest possible orbital distance. This could be done
by adding species transparent to the stellar irradiation but whose
combined absorption covers the whole mid-IR. The combined effect of
two or more species has to be studied in a self-consistent way. For
instance, stratospheric warming due to the NIR absorption of
CH$_4$ could induce more cooling than warming by preventing the formation
of CO$_2$ clouds. Doing the proper modeling, we may find that a planet
containing a cocktail of, say, CO$_2$, CH$_4$, N$_2$O, NH$_3$, and CFCs,
may extend surface habitability as far as 3~AU from the Sun (or 0.35~AU from Gl~581, see Sect.~\ref{sec:Gliese})
The problem with this approach is to justify how such a composition could be
sustained, taking into account the atmospheric sources and sinks, over
geological periods.  Life itself could be a possible answer. A planet
could be made habitable on purpose by an intelligent civilization able to
engineer the atmosphere.  ``Terraforming'' of Mars has for instance been
considered \citep{1991Natur.352..489M}. The \emph{greenhouse cocktail}
could also be maintained in a ``Gaia like'' homeostasis by some ecosystems
that have initially originated in the planet subsurface. We will leave
this tantalizing problem to further studies and will consider for the
moment that 2.4~AU is the outermost edge of the HZ for a Sun-like star.

\subsection{Limits of the HZ for non solar-type stars}\label{sec:HZstars}

\begin{figure*}
\begin{center}
\includegraphics[width=\linewidth]{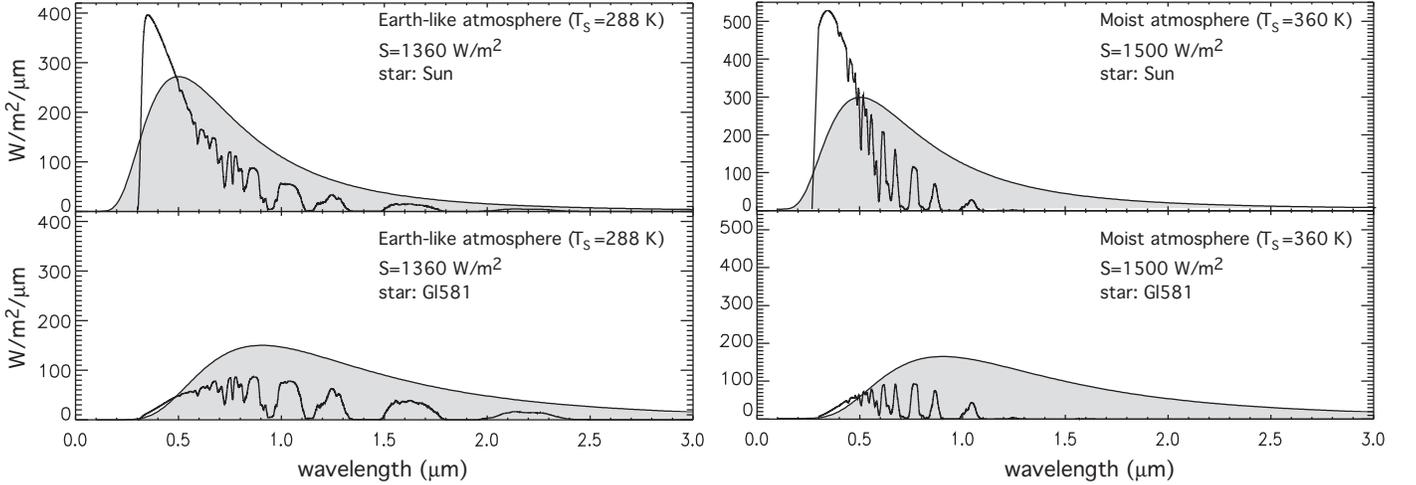}
\caption{Plots showing the effect of the stellar effective temperature on
the albedo. The graphs represent the reflected irradiance at the
substellar point of a planet subject to the irradiation from a Sun-like
star with $T_{\rm eff}=5800$~K (top) and from an M-type star with $T_{\rm
eff}=3200$~K, similar to Gl~581 (bottom). In this illustration, both stars
are assumed to radiate as black bodies. The graphs give the spectral
irradiance of a planet receiving an integrated stellar flux of
1360~W~m$^{-2}$ (left) and 1500~W~m$^{-2}$ (right). The shaded curves with
black-body shape give the irradiance of an airless planet with a fixed
surface albedo of 0.15. The non-shaded spectra show the irradiance
computed with Phoenix \citep{jimmythese} when a cloud-free atmosphere is
added. The current Earth atmosphere is shown on the left and a
N$_2$-H$_2$O atmosphere with $T_{\rm s}=360$~K, $P_{\rm s}$=1.3 bar and
$P_{\rm w}$=0.5 bar on the right. This illustrates the increase in the
reflected energy with the stellar effective temperature. }
\label{fig:albedo}
\end{center}
\end{figure*}

We have seen that the planetary albedo plays a crucial role in the
definition of the HZ. When the absorbed stellar flux per unit surface
S$\times$(1-A)/4 becomes higher than 300~W~m$^{-2}$, a runaway greenhouse
makes the planet uninhabitable. The albedo of the planetary surface is not
important when determining the inner and outer boundaries of the HZ
because the albedo of a habitable planet close to these edges would be
fully determined by its atmospheric composition (including clouds) and the
spectral distribution of the stellar radiation. The Sun emits a large
fraction of its energy in the visible, a wavelength domain where the
atmosphere of a habitable planet is highly reflective, because of the
dependence of Rayleigh backscattering as $\lambda^{-4}$ and because of the
lack of strong H$_2$O absorption bands. The emission of stars with a low
effective temperature peaks in the NIR (around 0.9~$\mu$m for Gl~581). In
the NIR, the contribution of Rayleigh backscattering to the albedo becomes
negligible and the strong absorption bands of H$_2$O (plus CO$_2$ and
possibly CH$_4$ in the outer HZ) cause additional absorption of stellar
radiation. This effect is illustrated on Fig.~\ref{fig:albedo}.

Because of the relation between the albedo and the effective temperature
of the star, the limits of the HZ cannot be simply scaled to the stellar
luminosity. For stellar effective temperatures between 3700 and 7200~K,
Kasting et al. \citeyearpar{1993Icar..101..108K} calculated the albedo of
a planet near both edges of the HZ that has either a dense H$_2$O or a
dense CO$_2$ atmosphere. Around an M-type star with $T_{\rm eff}=3700~K$,
for instance, accounting for this effect represents a 40\% difference in
the stellar flux (15\% in orbital distance) for a cloud-free H$_2$O-rich
atmosphere close to the inner edge. The scaling factor for a planet with a
thin atmosphere, like the modern Earth, is smaller -- about 10\% in terms
of stellar flux, or 5\% in semi-major axis \citep{2005AsBio...5..706S}.  
From the various cases studied by Kasting et al. (1993), the limit $l_{\rm
in}$ and $l_{\rm out}$ of the solar HZ can be extrapolated to any star
with luminosity $L$ and an effective temperature between 3700~K and 7200~K
by using the following relationships:  \begin{equation} l_{\rm in} = \left
( l_{\rm in \odot } - a_{\rm in} T_{\star} - b_{\rm in} T_{\star} ^2
\right ) \left ( \frac{L}{L_{\odot}} \right)^\frac{1}{2} \label{eq:fit1}
\end{equation} \begin{equation} l_{\rm out} = \left ( l_{\rm out \odot } -
a_{\rm out} T_{\star} - b_{\rm out} T_{\star} ^2 \right ) \left (
\frac{L}{L_{\odot}} \right)^\frac{1}{2} \label{eq:fit2} \end{equation}
with $a_{\rm in}=2.7619\times10^{-5}$, $b_{\rm in}=3.8095\times10^{-9}$,
$a_{\rm out}=1.3786\times10^{-4}$, $b_{\rm out}=1.4286\times10^{-9}$, and
$T_{\star} = T_{\rm eff}-5700$. Here, $l_{\rm in}$ and $l_{\rm out}$ are in AU,
and $T_{\rm eff}$ in K.

As discussed above, the values of $l_{\rm in \odot}$ and $l_{\rm out
\odot}$ depend on the criteria chosen to define the limit of habitability.
Table~\ref{tab:limits} gives the limits of the present solar HZ based on
the ``early Mars'' and ``recent Venus'' criteria, and the
radiative-convective models with a cloudiness of 0, 50, and 100\%.  For
each cloud coverage, two $l_{\rm in \odot}$ values are given: the runaway
greenhouse limit, and the $T_{\rm s}=373$~K limit.

\begin{table}[t]
\caption{Boundaries of the present Solar HZ}
\begin{center}
\begin{tabular}{lllll}
  \hline
& Venus & clouds & clouds & clouds \\
&crit.& 0\% & 50\% & 100\%  \\
\hline
$l_{\rm in \odot}$ (AU)  & 0.72 & 0.84-0.95 & 0.68-0.76 & 0.46-0.51\\
\hline
& Mars & clouds & clouds & clouds  \\
&crit.&  0\% & 50\% & 100\%  \\
\hline
$l_{\rm out \odot}$ (AU) & 1.77 & 1.67 & 1.95 & 2.4 \\
\hline
\end{tabular}
\end{center}
\label{tab:limits}
\end{table}

Note that the effect of the spectral type on the albedo, included in Eqs.
(\ref{eq:fit1}) and (\ref{eq:fit2}) as a quadratic function of ($T_{\rm
eff}-5700$), was estimated only for a cloud-free atmosphere. Since the
reflectivity of clouds is less sensitive to wavelength, this quadratic
term may not be valid to scale the boundaries of the HZ for planets
covered by clouds. The actual boundary locations can be bracketed by using
the simple luminosity scaling and Eqs. (\ref{eq:fit1}) or (\ref{eq:fit2})
that include the $T_{\rm eff}$ sensitivity.

\subsubsection{The continuously habitable zone}

\begin{figure}[t]
\begin{center}
\includegraphics[width=\linewidth]{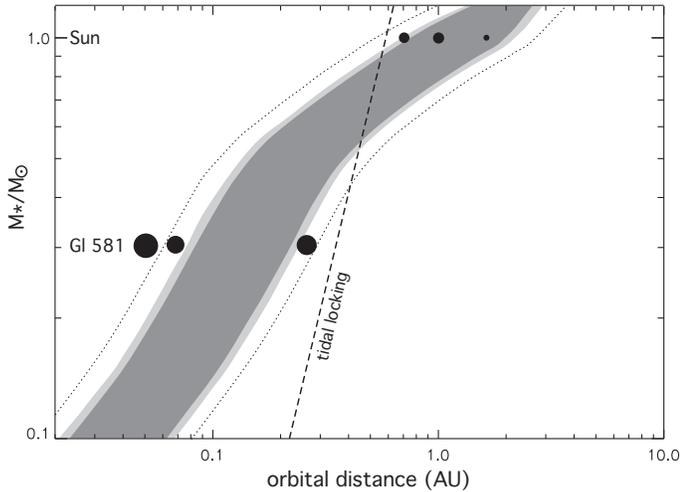}
\caption{The orbital region that remains continuously habitable during at
least 5 Gyr as a function of the stellar mass. The darker area is
defined by the empirical ``early Mars'' and ``recent Venus'' criteria. The
light grey region gives the theoretical inner (runaway greenhouse) and
outer limits with 50\% cloudiness, with H$_2$O and CO$_2$ clouds,
respectively. The dotted boundaries correspond to the extreme theoretical
limits, found with a 100\% cloud cover. The dashed line indicates the
distance at which a 1~$M_{\oplus}$ planet on a circular orbit becomes
tidally locked in less than 1~Gyr.}
\label{fig:CHZ}
\end{center}
\end{figure}

The luminosity of a star, and thus the boundaries of its HZ, change during
its lifetime. Therefore, another criterion for planetary habitability
could be the time spent at a habitable distance. For any star of a given
spectral type, it is possible to define the limits of the region that
remains within the HZ for a time period longer than a selected timespan.
Figure~\ref{fig:CHZ} shows the boundaries of the continuously HZ (CHZ) as a
function of the stellar mass, computed with Eqs.~(\ref{eq:fit1}) and
(\ref{eq:fit2}) and with an evolutionary model for stars of solar
metallicity \citep{1998baraffe}. Since Eqs.~(\ref{eq:fit1}) and
(\ref{eq:fit2}) are not valid for stellar $T_{\rm eff}$ below 3700~K, the
effect on the albedo was calculated by assuming $T_{\rm eff}=3700$~K for
temperatures below this value.

\section{The cases of Gliese 581c and Gliese 581d}\label{sec:Gliese}

\begin{figure*}[t]
\begin{center}
\includegraphics[width=0.8\linewidth]{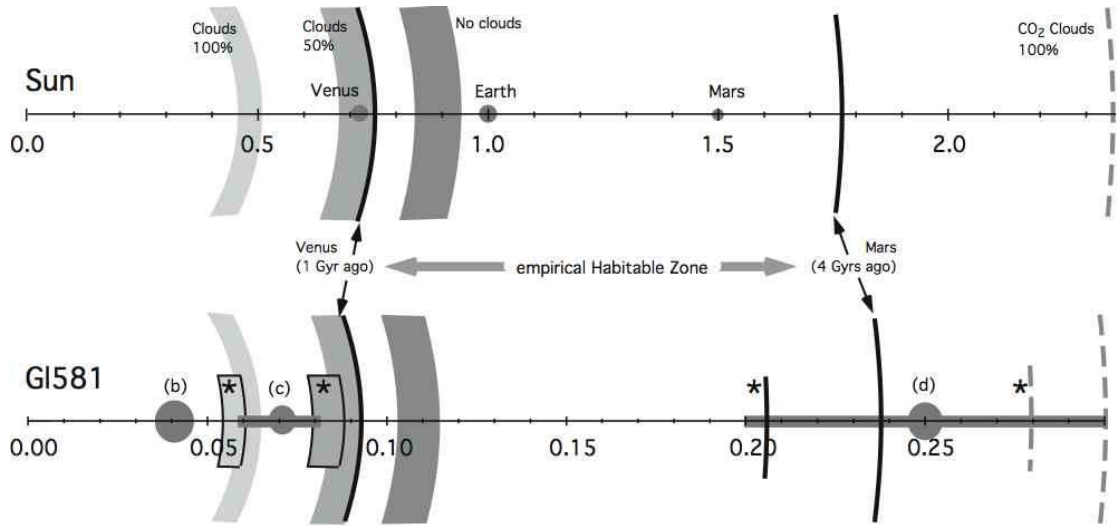}
\caption{Diagrams depicting the HZ around the Sun and Gliese
581. The grey areas indicate the theoretical inner edge for different
fractional cloud covers. The width of each inner edge is defined by the
runaway greenhouse and water loss limits. The thick lines give the
inner and outer boundaries of the ``empirical'' HZ, based on
the non-habitability of Venus for the last 1 Gyr and the apparent
habitability of Mars 4~Gyr ago. The dashed line gives the outermost
theoretical limit of the HZ, found with a 100\% CO$_2$ cloud cover. The
upper diagram shows the limits computed for the Sun's properties. The
lower diagram shows the limits computed for a 3700~K M-type star and
scaled to the stellar luminosity of Gl~581 (for which $T_{\rm
eff}=3200$~K). The limits marked with the symbol * are simply scaled to
the luminosity and are thus slightly closer to the star -- they are shown
here because the wavelength dependency of the albedo has been explicitly
computed only for thick and cloud-free H$_2$O- and CO$_2$-rich
atmospheres. The albedo should be less sensitive to the stellar effective
temperature in the presence of clouds. For the eccentric solution found by
Udry et al. (2007), the orbits are confined within the grey bars about the
marked planet locations.}
\label{fig:HZ}
\end{center}
\end{figure*}

The star Gl~581 has an inferred effective temperature of 3200~K and a
luminosity of 0.013~L$_\odot$ \citep{2007A&A...469L..43U}.
Equations~(\ref{eq:fit1}) and (\ref{eq:fit2}) cannot be applied for
$T_{\rm eff}<3700$~K because no radiative-convective simulation has been
performed for such low stellar effective temperatures. However, the albedo
calculated for $T_{\rm eff}=3700$~K and 3200~K should be similar.
Figure~\ref{fig:HZ} shows the location of the boundaries of the HZ around
Gl~581 for various criteria, along with the three known planets. The
limits are given with or without the albedo correction (except for the
cloud-free simulation, which was specifically run for $T_{\rm
eff}=3700$~K).

\subsection{The planet Gl~581c}

Figure~\ref{fig:HZ} shows that planet Gl~581c would be habitable only if
clouds with the highest reflectivity covered most of the daytime
hemisphere.  A 50\% cloud cover is not enough to prevent a runaway
greenhouse effect on Gl 581c, which receives 30\% more energy than Venus
today. This problem is exacerbated by the fact that Venus has a much
higher albedo than the expected value for a habitable planet at the
orbital distance of Gl~581c. The composition of the atmosphere of Gl~581c
depends on the mass of the initial water reservoir on the planet, and on
the efficiency of the gravitational escape of H. Two possible scenarios,
inspired by the fate of Venus and by simulations done for an Earth-like
planets, can be suggested:
\begin{itemize}
\item[$i)$] The reservoir of water is large so only a fraction of it will
be lost to space. Numerical simulations suggest that planets more massive
than about 5~$M_\oplus$, such as those found around Gl~581 or HD~69830
\citep{2006A&A...455L..25A}, started their formation in the cold outer
protoplanetary disk, accreting some icy planetesimals, and migrated close
to the star. In this situation, water can be orders of magnitude more
abundant than on Earth. If this is the case for Gl~581c, its rocky surface
is probably covered by a thick layer of H$_2$O. Depending on the amount,
water can form a mantle of hot and high-pressure ice underneath a fluid
envelope of supercritical H$_2$O (Selsis et al. 2007).
\item[$ii)$] The reservoir of H$_2$O has been lost earlier, when the star was
still a strong X-ray and EUV emitter. A CO$_2$-rich, Venus-like atmosphere
could remain, with surface temperatures still in the range 700--1000~K. A
significant level of leftover O$_2$ may be present.
\end{itemize}
But the planet could also have a bulk composition that is so different from Earth
that a broader diversity of situations should be explored. The planet
could, for instance, be a volatile-rich planet
\citep{2003ApJ...596L.105K,2004ocean_planets} or a {\em carbon}- or
{\em ceramic}-planet \citep{2005astro.ph..4214K,2000Icar..145..637G},
with an atmospheric composition very different from our assumptions
derived from the study of Venus, Earth, and Mars. However, water is
considered an indispensable prerequisite for life and any additional
atmospheric constituent would likely enhance the opacity of the atmosphere
and the amount of greenhouse warming. The inner edge determined for a pure
H$_2$O atmosphere thus seems to be the closest possible limit for
habitability.

\subsubsection{The effect of gravity}

The gravity at the surface of Gl~581c, calculated for its minimum mass and
for the range of radii predicted by Sotin et al. (2007) or Valencia et al.
(2007), is between 1.3 and 2~$g$. Therefore, a given value of the vapor
pressure $P_{\rm w}(T_{\rm s})$ would correspond to a vapor column
1.3--2.0 times smaller than the equivalent at Earth gravity. As the water
column determines the IR opacity of the atmosphere, a higher gravity is
expected to shift the inner edge of the HZ closer to the star. However, IR
opacity is not the only parameter to be affected by the gravity. For a
given value of $T_{\rm s}$, gravity also influences the lapse rate and the
albedo produced by the water column. When treated self-consistently, these
effects tend to compensate for each other, thus weakening the overall
influence of gravity. Radiative-convective simulations of a planet with a
surface gravity of 2.5~$g$ were carried out by Kasting et al.
\citeyearpar{1993Icar..101..108K}, who found an inner edge only 3\%
closer. For the expected gravity of Gl~581c, its effect can be safely
neglected.

\subsubsection{Stability of a quasi-snowball state}

For a given stellar irradiation, there is not necessarily a unique
solution for $T_{\rm s}$. On Earth, for instance, climate is known to be
bistable. It has been warm during most of its history, with a mean surface
temperature well above 273~K (which includes glacial and interglacial
periods), except for a few {\em snowball Earth} events characterized by
an ice cover down to the equator. The most recent snowball events occurred at
the end of Neoproterozoic era between 730 and 610~Myr ago. At the
beginning of these events, the runaway ice-albedo feedback makes the
global mean temperature drop to $\sim$220~K for a few tens of thousands of
years. This temperature drop is followed by a period of a few million
years with a mean temperature around 265~K \citep{2002GGG....3fR...1S}.
Thanks to the volcanic release of CO$_2$ and the inefficiency of carbonate
formation on a frozen Earth, our planet did not remain trapped in such
state. This constitutes strong evidence of the long-term stabilization of
the climate through the carbonate-silicate cycle. Snowball events
illustrate the fact that the Earth would not be close enough to the Sun to
maintain $T_{\rm s}$ above 273~K if water vapor was the unique greenhouse
gas in the atmosphere. Without enough atmospheric CO$_2$, the surface of
our planet would be frozen, and the very high albedo of the frozen oceans
would permanently keep $T_{\rm s}$ as low as 220~K.

It can be considered whether such bistability could be possible on
Gl~581c. Radiative-convective models give one solution for the atmosphere
structure and surface temperature, but would a {\em cold solution} be
possible? Let us assume that ice covers the entire planetary surface,
resulting in an albedo of $\sim$0.8, and that the atmosphere has
a negligible radiative effect, either because it is very tenuous or because
it is composed of IR-transparent gases like N$_2$. In this case, $T_{\rm
eq}$ would be close to 235~K, which is consistent with this
``snowball'' hypothesis. However, in the absence of a greenhouse effect,
and in the case of a slowly rotating planet (see Sect.~\ref{sec:sync}),
the surface temperature at the substellar point would be 330~K. This
temperature corresponds to a vapor pressure of 0.2~bar, which is
inconsistent with the assumed albedo and atmospheric transparency. If we
assume an extreme albedo of 0.95, as on Enceladus, the icy satellite of
Saturn, the substellar temperature would be well below 273~K, and the icy
state would be stable. However, in the absence of liquid water on the
surface, the planet would not be habitable. An interesting situation might
be found for a very narrow range of albedo ($A=0.89-0.90$) for which most
of the planetary surface is frozen, except for an area at the substellar
point, where the local equilibrium temperature and the vapor pressure
slightly exceed 273~K and 6.1~mbar, respectively. This would allow for the
presence of liquid water.

In this latter case, though, the IR opacity and effect on the albedo of a
6.1 mbar vapor column is already important (in fact the mean water content
of Earth's atmosphere is very close to this value) and is not consistent
with the assumed surface temperature. This means that temperature and
vapor pressure are expected to diverge rapidly from this assumed starting
point. It would be interesting to study this case with a time-dependent
model including the vaporization of water at the substellar point and its
condensation in the cold regions, to check whether a substellar hot spot
can exist without triggering a runaway greenhouse. Such a model should also
include the rotation of the planet and the variation in the orbital
distance due to the eccentricity. Indeed, a partially habitable
steady-state implies that the substellar point moves on the planetary
surface (otherwise all the water inventory ends up in the frozen regions),
which requires significant eccentricity (see Sect.~\ref{sec:sync}).
However, this case is {\em ad hoc} as there is no particular reason why
the planetary albedo would have this exact value and, if the planet was
mainly frozen, any volcanic release of CO$_2$ (expected on a planet as
massive as Gl~581c) would trigger a runaway greenhouse.

\subsection{The planet Gl~581d}

Planet Gl~581d receives about half the energy flux that Mars gets currently from the Sun. 
However, because CO$_2$-rich atmospheres absorb more energy from
an M-type star than from a G-type star, the orbital distance of Gl~581c is
no more than 4\% beyond the empirical ``early Mars'' limit (when scaled to
Gl~581). Moreover, a partial cover of CO$_2$-ice clouds can theoretically
sustain habitable conditions for even lower stellar fluxes. The third
planet of this system is thus potentially habitable, according to our
present knowledge. Its high mass makes the maintenance of a thick
atmosphere possible over billions of years. Of course, the high
($\sim8$~$M_{\oplus}$) minimum mass also means that the actual mass could
be $>10$~$M_{\oplus}$, and the planet may thus be a gas or ice giant,
rather than a rocky planet. If so, this planet is only one of the many
other gas giant exoplanets already known to be within the HZs of their
parent stars.

The warming effect of CO$_2$-ice clouds has only been studied for Sun-like
irradiation. Because of their size and their optical constants, typical
CO$_2$-ice particles are relatively transparent to visible radiation, but
scatter efficiently at mid-IR wavelengths. For a stellar flux shifted
towards the NIR, the albedo of these clouds could be significantly higher
than for solar irradiation. This effect is not included in
Eq.~(\ref{eq:fit2}). Detailed modeling is thus required, without which it
is not possible to determine the precise location of the edge of the HZ as
defined from CO$_2$-ice clouds, and thus the potential habitability of the
large terrestrial planet Gl~581d.

As discussed in Sect.~\ref{sec:super-greenhouse}, some atmospheric
compositions with the right {\em greenhouse cocktail} might provide 
stronger greenhouse warming than CO$_2$ alone and sustain liquid water at
the surface of Gl~581d. In addition, some greenhouse gases, such as
CH$_4$, condense at much lower temperatures than CO$_2$ (the triple point
of CH$_4$ is at 90~K and 0.1~bar) and would not be trapped as ice on the
night side in case of synchronous rotation (see Sect.~\ref{sec:sync}). It
is not obvious that CH$_4$ alone could maintain habitable conditions on
Gl~581d. Indeed, the efficiency of the surface warming is expected to
decrease above a certain CH$_4$ level. This is due to the stratospheric
heating by the direct absorption of the visible and near-IR stellar flux.
For Sun-like irradiation and partial pressures of CH$_4$ above about
5~mbars, this effect competes with the surface greenhouse warming. The
temperature structure of a CH$_4$-rich atmosphere under the irradiation of
the star Gl~581 should thus be specifically calculated to answer this
question. A remaining problem is to identify the geochemical mechanisms
able to stabilize a CH$_4$-rich or a ``super-greenhouse'' atmosphere over
several Gyr.

\subsection{An excess of volatiles?}\label{sec:volatiles}

The volatiles that constitute the atmosphere of terrestrial planets in the
Solar System have been mainly accreted as solids (rocks, with a possible
minor contribution from ices). This is inferred from the composition in
noble gases and their isotopes \citep{1991Icar...92....2P}. If a
significant mass of H$_2$ and He was accreted directly from the
protoplanetary nebula, most of it was lost to space and the traces that
remain represent a negligible fraction of the present atmospheres. The
last phases of the accretion of Earth are likely to have occurred after
the dissipation of the protosolar nebula. Indeed, disks are found to 
typically live less than 5--10~Myr \citep{2007prpl.conf..573M}, while it should
have taken more than about 30 Myr to form the Earth
\citep{2006EM&P...98...97M}. However, due to the diversity of disk
properties (in lifetime, mass, density) and the role of migration (which
accelerates the accretion), it is possible that the accretion of nebular
gas is much more efficient on some terrestrial proto-exoplanets,
especially if they are more massive than Earth. Rafikov
\citeyearpar{2006ApJ...648..666R} found that planets more massive than
6~$M_{\oplus}$ could have accreted a significant fraction of their mass as
H$_2$-He. For instance, a 8~$M_{\oplus}$ planet could be made of
7~$M_{\oplus}$ of rocks and 1~$M_{\oplus}$ of H$_2$-He. The dissipation of
the disk could have frustrated the evolution of such planet into a gas
giant. It is thus possible that some planets more massive than
$\sim$6~$M_{\oplus}$ (which is potentially the case of Gl~581c and Gl
581~d) could not host liquid water because of the pressure and temperature
imposed by a massive envelope of gas. This situation would be similar to
that of HD 69830d.  This 18~$M_{\oplus}$ planet found in the HZ of its
star \citep{2006Natur.441..305L} is thought to have accreted a massive
H$_2$-He envelope, underneath which water can only exist as a
supercritical fluid or high-pressure ice \citep{2006A&A...455L..25A}.

As mentioned before, the Gl~581 planets may very well have started their
formation beyond the snow line. In this case, and even without invoking
the accretion of hydrogen-rich gas from the protoplanetary nebula, their
volatile content could be orders of magnitude higher than on Earth. If
only 10\% of the accreted solids are made of cometary ice, a
6~$M_{\oplus}$ planet would contain as much as 0.55~$M_{\oplus}$ of water
and 0.06~$M_{\oplus}$ of other volatiles, mainly CO$_2$, CO, CH$_4$, and
NH$_3$. For comparison the mass of the terrestrial oceans is
$2\times10^{-4}$~$M_{\oplus}$ and the mass of Earth's atmosphere is
$9\times10^{-7}$~$M_{\oplus}$.

The distribution of these volatiles between the interior and the
atmosphere (including here surface ices) at the end of the accretion
depends on the planet's thermal history.  Also, the volatiles
initially present in the planetesimals may not be retained during the
violent accretional collisions \citep{2007ApJ...660L.149L}. However, the
planet can possibly have a massive and enriched gaseous envelope
preventing the liquid phase of water from existing on the surface.

\subsection{Atmospheric erosion and the habitability of planets around
M-type stars} \label{sec:erosion}

Models presented in this paper assume that planets can retain a dense atmosphere. However, planets in the HZ of active stars can be exposed to high levels of X-ray
and EUV radiation (XUV) and strong particle fluxes from the quiescent
stellar wind or coronal mass ejections (CMEs). Such high emissions are the
result of the stellar magnetic activity and can induce important thermal
and non-thermal atmospheric losses to space, potentially able to strip the whole atmosphere
 \citep{2005AsBio...5..587G,2006SSRv..122..189L,2007AsBio...7...27L,2007AsBio...7..167K,2007AsBio...7..185L,2007AsBio...7...85S}.
Within the HZ of the Sun and solar-type stars, conditions threatening the survival of the atmosphere and the habitability are limited to the first few hundred million years
\citep{2007SSRv..129..207K}, and we know that the Earth and Venus atmospheres survived this early active phase (more damage may have been caused to Mars). But extreme irradiation conditions could last several Gyr in the case
of M-type stars, as discussed below. In the context of Gl 581, it is thus
worth addressing the question of stellar activity and the impact on the
habitability of its planets.

For an initial estimate of the evolution of XUV irradiances we have used a
proxy indicator, which is the ratio of the X-ray luminosity to the
bolometric luminosity ($\log \left[L_{\rm X}/L_{\rm bol}\right]$). This
ratio is highest for the most active stars (i.e. fastest rotation period)
and decreases monotonically with decreasing level of chromospheric
activity (e.g., Stelzer \& Neuh\"auser, \citeyear{2001A&A...377..538S};
Pizzolato et al. \citeyear{2003A&A...397..147P}). From
the analysis of open-cluster stars it is now well established that all
single late-type stars (G-K-M) spin down as they age. Their activity
decreases with time, and so does the ratio $\log (L_{\rm X}/L_{\rm bol})$.
It is also a well-known effect that $\log (L_{\rm X}/L_{\rm bol})$ does
not increase up to values arbitrarily close to unity for very active
stars. Instead, a ``saturation'' phenomenon occurs and no active star
seems to go higher (except for the strongest flares) than $\log (L_{\rm
X}/L_{\rm bol})\approx -3$ (e.g., Vilhu \& Walter, \citeyear{1987ApJ...321..958V}; Stauffer et al.,
\citeyear{1994ApJS...91..625S}). Qualitatively, the evolution of $\log (L_{\rm X}/L_{\rm bol})$ for
a late-type star has a flat plateau from its arrival on the main sequence
up to a certain age (end of saturation phase) and then decreases
monotonically as a power law function of age.

\begin{figure}[t]
\begin{center}
\includegraphics[width=\linewidth]{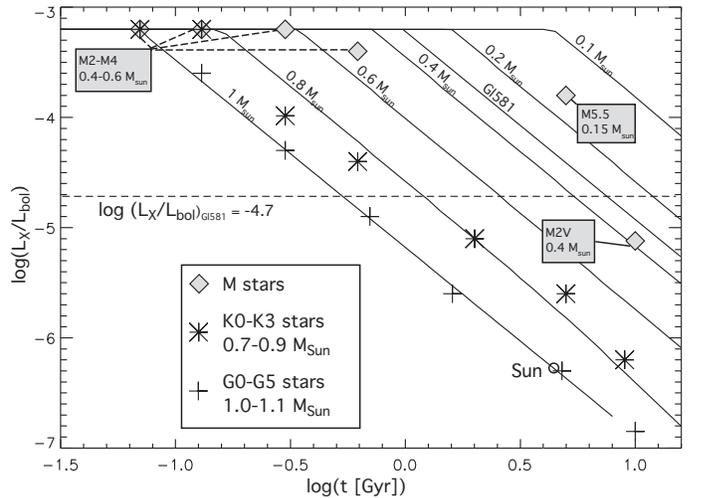}
\caption{Evolution of the ratio between the X-ray and
bolometric luminosities as a function of age for stars of different
masses. The solid lines represent semi-empirical laws, while symbols give
observed values for G (+), K ($*$) and M ($\diamond$) stars. The dashed
line gives the upper limit for the value of $\log (L_{\rm X}/L_{\rm bol})$
in the case of Gl~581, for which there is no ROSAT detection.}
\label{fig:Lx_Lbol}
\end{center}
\end{figure}

We have compiled a sample of K- and M-type stars with ages determined
in a similar way to the solar analogs in the ``Sun in Time'' sample (Ribas et al.
2005), i.e., membership in clusters and moving groups \citep{2001MNRAS.328...45M}, rotation period, membership in wide binaries, and isochrones. For
these, $\log (L_{\rm X}/L_{\rm bol})$ values have been obtained from the
thorough list provided by Pizzolato et al. (2003) and complemented in a
few cases with values estimated directly from ROSAT measurements following
Schmitt et al. \citeyearpar{1995ApJ...450..392S}. The evolution of $\log (L_{\rm X}/L_{\rm bol})$
with age for stars of different spectral types is illustrated in Fig.
\ref{fig:Lx_Lbol}.  Note that today's Sun has a value of $\log(L_{\rm
X}/L_{\rm bol}) \sim -6.1$.  The solid lines are semi-empirical estimates
that result from the fact that the evolution of $L_{\rm X}$ in the power
law regime is roughly independent of the spectral type and $L_{\rm bol}$
comes from the use of stellar models for different masses. For the K- and
M-type stars these are preliminary estimates and the uncertainty of each
point can be of a few tenths of a dex. In the case of G-type stars, the
values plotted are much more reliable and come from direct measurements
for the ``Sun in Time'' targets. More details on this analysis will be
given in a forthcoming publication (Ribas et al. 2008, in prep.).

It is reasonable to assume that the overall XUV flux received by a planet in
the HZ scales with the ratio $\log (L_{\rm XUV}/L_{\rm bol})$. Figure
\ref{fig:Lx_Lbol} shows that, while solar-type stars stay at saturated
emission levels until ages of $\sim$100 Myr and then their XUV
luminosities rapidly decrease following a power-law relationship as a
function of age, M-type stars have saturated emissions (i.e., highest
activity) up to ayes of a few Gyr. Thus, planets in the HZ of M-type stars 
may receive XUV fluxes that are 10--100 times higher than in the HZ of solar-type stars of the
same age.

To estimate the possible activity level of Gl 581 we carried out a search 
in the ROSAT Faint Source Catalog \citep{2000IAUC.7432R...1V} that produced no
result. Thus, Gl 581's X-ray flux is below the ROSAT detection threshold. 
Such a threshold can be calculated following Schmitt et al. (1995) and we found 
that $L_{\rm X}$ must be lower than $10^{27}$~erg~s$^{-1}$ at the distance of Gl 581, implying 
that $\log(L_{\rm X}/L_{\rm bol}) < -4.7$.  From Fig. \ref{fig:Lx_Lbol}, 
this leads to a lower limit of the age that, considering the associated 
uncertainties, could be around 7 Gyr. However, we caution that this is 
still a preliminary estimate and that Gl 581 should be studied further 
before giving a conclusive figure. With regards to an upper 
limit to the age, the space motions of Gl 581 
$(U,V,W)=(-22.7,-29.3,-8.0)$~km~s$^{-1}$ are characteristic of a galactic 
disk star, and thus its age should not be above 10--11 Gyr. Also, the star 
is known to have a only midly subsolar metallicity ($[Fe/H\approx-0.3$; 
Bonfils et al. \citeyear{2005A&A...443L..15B}; Bean et al. \citeyear{2006ApJ...653L..65B}), 
which is also consistent with being a disk member.

The current XUV emissions of Gl~581 may not represent a significant threat
to the stability of planetary atmospheres inside the HZ, but these planets
have most likely suffered strong erosion phenomena during earlier stages.
Dominant loss mechanisms are thermal escape, induced by the heating of the
upper atmospheric layers by XUV irradiation, and non-thermal escape caused
by the interaction of these upper layers with the stellar particle wind.
The efficiency of these erosion processes depends on several factors.
First, the composition of the atmosphere controls the temperature of the
exosphere and thus the thermal escape rates, which can be extremely high
if the blow-off temperature is reached \citep{2006SSRv..122..189L}. A high
mixing ratio of CO$_2$, for instance, allows the exosphere to cool
efficiently by IR emission (that is why the exosphere of Venus is about
three times cooler than Earth's). The thermal structure of the upper
atmosphere also determines the extension of the exosphere and indirectly
the efficiency of non-thermal losses (like sputtering, ion-pick up,
dissociative recombination). Non-thermal losses also depend on poorly
constrained parameters like the intrinsic magnetic moment of the planet
and on the stellar wind properties (velocity and density).

Observational constraints on stellar winds exist
\citep{2005ApJ...628L.143W}, but their interpretation is still debated
\citep{2007A&A...463...11H}. It does however seem well established that
the particle winds of active M-type stars could at least be 10 times
stronger than that of today's Sun, in the HZ. During the early and active stages of
an M-type star like Gl~581, the rate of atmospheric losses due to the
association of XUV heating and particle wind-induced ion pick-up could
theoretically reach tens of bars of heavy elements (O, C, N) per Gyr for a
weakly-magnetized planet in the HZ, even with a CO$_2$-rich atmosphere
\citep{2007AsBio...7..185L}. Tidal interactions rapidly slow down the
rotation of planets in the HZ of M-type stars (see Sect.~\ref{sec:sync}),
which may result in weak intrinsic magnetic moments, unable to provide
sufficient protection against non-thermal erosion.  On the other hand,
Gl~581c and Gl~581d are significantly more massive than the Earth, which
may result in an enhanced dynamo and better shielding. Their surface
gravities are also higher by factors 1.3 and 2.5, respectively, which
limits the extension of the exosphere and should lower the erosion rate.

Eventually, the threat posed by the atmospheric erosion depends on the
inventory of volatiles, as has already been discussed in the case of the water
loss in Sect.~\ref{sec:waterloss}. If the loss rate exceeds the outgassing
rate, or if the atmospheric losses occurring during the planet's early
stages exhaust species essential for the maintenance of life, the planet
would obviously not remain habitable in the long term. Gaidos \& Selsis
\citeyearpar{2007prpl.conf..929G} point out that a replica of early Earth
(assumed to have a CO$_2$-rich atmosphere) located in the HZ of a M-type
star could lose its entire atmospheric nitrogen inventory during its first
Gyr via thermal escape from XUV heating alone. A large reservoir of
volatiles and/or a continuous outgassing of the essential atmospheric
species is thus required to ensure durable habitable conditions around
M-type stars. 

Models providing quantitative estimates for atmospheric escape in 
the extreme conditions encountered in the HZ of M stars are still under 
development. Available preliminary results cited in this section are 
extremely promising and stress the importance of atmospheric retention for 
habitability around low-mass stars. However, current estimates of thermal 
and non-thermal losses in this context assume a static upper atmosphere, 
while high atmospheric escape rates would require a hydrodynamic 
treatment. This field of research has recently become very active due to 
the important questions raised by the habitability of planets around 
M-type stars. Although we have to end this section with a question mark, 
the reader should expect forthcoming results obtained with thermal 
hydrodynamic models coupled with non-thermal erosion that will shed more light 
on the fate of planetary atmospheres around active stars.

\subsection{The effect of rotation and eccentricity}\label{sec:sync}

Udry et al. (2007) present two fits of the radial velocity curve, one
obtained with fixed circular orbits, and another where the eccentricity is
a free parameter. The latter resulted in the best agreement and yielded
$e=0.16$ and $e=0.2$ for Gl~581c and Gl~581d, respectively.

If the orbits are circular, then these planets could have evolved into a
synchronized state, in which the rotation period ($P_{\rm r}$) equals the
orbital period ($P$). The synchronization time of these planets is indeed
very short. We obtain 25\,000~yr and $10^{7}$~yr for planets Gl~581c and
Gl~581d, respectively, when adopting a value of $Q=100$ for the
dissipation factor, typical of terrestrial planets. In this case, the
planets have permanent day and night hemispheres. It is interesting to
note that we find a much longer circularization time for these two planets
($10^{10}$~yr), which is consistent with eccentric orbits. If the planets
do have a significant eccentricity, then the following cases are possible:
\begin{itemize}
\item[$i)$] Because terrestrial planets are expected to have a non purely
axisymmetric shape, tidal spin down may have made it possible for the
rotation to be captured in a spin-orbit resonance with $P=(n/2) P_r$.
Mercury, for instance, is trapped in a 3:2 spin-orbit resonance as a
consequence of its oblong shape. For resonances with $n \neq 2$, the star
rises and sets at any point of the planet and there is no permanent dark
area. In the synchronous case ($n=2$), there is a region of the planet
that never receives direct starlight. For an eccentric planet with zero
obliquity, the size of the corresponding surface is governed by the
amplitude of the optical librations about the substellar point when
neglecting the smaller contribution of forced and free librations.
Considering up to third-order terms in eccentricity, we estimate
that the amplitude of the libration is $2\,e+25\,e^3/16$, and thus the
fractional area $r$ of the planet that can receive direct starlight, is
then given by
\begin{equation}
r=\frac{1}{2}+\frac{2\,e+25\,e^3/16}{\pi}.
\end{equation}
For $e=0.2$, the optical libration amounts to $\sim 23^{\circ}$, and $r$
would take a value of 63\%. Any non-zero obliquity adds latitudinal
librations that may significantly increase these values. For high
obliquities, it is possible for the entire surface of the planet to
receive direct starlight some time during an orbital period. However, high
obliquities are not consistent with tidally-driven evolution models of the
spin axis.
\item[$ii)$] The planet has not been captured in a spin-orbit resonance. In
this case, no region of the planet is permanently dark.
\end{itemize}

Let us now consider the synchronized case of $P_{\rm r}=P$, with low
obliquity. Whether the orbit is eccentric or not, there is a region on the
planet that is not directly heated by the stellar radiation. Such a region
can be an entire hemisphere if the eccentricity is negligible. There is
thus a risk of the atmospheric constituent condensing irreversibly on an
ice sheet covering the dark region.  Joshi et al. (1998) and Joshi (2003)
have shown that the atmosphere has to be sufficiently dense and opaque in
the IR to transport enough heat to the dark area and to prevent such an
atmospheric collapse. Near the edges of the HZ, a habitable planet
fulfills these conditions because it is very rich in either H$_2$O or
CO$_2$. Therefore, if one concludes that planets Gl~581c and Gl~581d are
habitable thanks to the effect of clouds, then synchronized rotation would
not appear to represent an impossible obstacle to their habitability
(although this would give rise to a climate that is certainly not ``Earth
like'', with a {\em super-rotating} atmosphere). In addition, the
strong irradiation of planet Gl~581c is expected to produce strong thermal
tides in the atmosphere (similar to those on Venus) that could prevent the
synchronization of the planet \citep{2003Icar..163...24C}.

For planet Gl~581d, whose orbit is in the outer area of the HZ,
synchronization can pose a more significant threat. The habitability of
the surface of the Earth has remained constant during most of its history,
except for a few snowball events possibly generated by climatic
instabilities. The Earth was able to recover from these frozen episodes
thanks to the continuous release of volcanic CO$_2$. However, a
synchronized planet may not be able to recover from a snowball event as
effectively as the Earth did because the condensation of CO$_2$ on the dark
side would occur much faster than the release of volcanic CO$_2$. Only a
rapid and dramatic event, such as a large impact on the ice sheet, could
initiate a new habitable period. The habitable state of a synchronized
planet could thus be particularly fragile when it relies on the presence
of atmospheric CO$_2$, which is the case for most of the HZ except for the
very narrow inner part where H$_2$O is ``self sufficient''.

Long-term habitability on a synchronized planet could still be maintained
if the greenhouse warming was provided by atmospheric compounds that
remain gaseous at sufficiently low temperature, such as CH$_4$. In this
situation, one may wonder if an inefficient transport of the incident
energy from the starlit to the dark hemisphere could help maintain
habitable conditions on the starlit hemisphere, or in a smaller region
around the substellar point, while the dark hemisphere would remain too
cold to host liquid water. In this case, restricting the habitability to a
fraction of the planetary surface would compensate for the low stellar
flux. Although this case represents a tantalizing configuration, the cold
trap for water would still operate, and this would be irreversibly
transported from the habitable to the frozen region.

In the case of the planets having an eccentricity as high as 0.16--0.2, the
study of the climate at steady state is of course less relevant than for a
circular orbit. As shown by Fig.~\ref{fig:HZ}, both Gl~581c and Gl~581d
are likely to make incursions inside the HZ (or excursions outside the HZ)
depending on the assumed boundaries. This situation is not expected to
change the discussion significantly regarding planet Gl~581c, since the
timescales associated with the vaporization of the water reservoir are
significantly longer than the 12-d orbital period. In addition, Williams
\& Pollard \citeyearpar{2002IJAsB...1...61W} show that habitability
depends primarily on the average stellar flux received over an entire
orbit, $<S>$, even at high eccentricity, and $<S>$ depends on the eccentricity
through the relationship:
\begin{equation}
<S> = \frac{S}{\sqrt{ 1 - e^{2}}}
\end{equation}
where $S$ is the stellar flux at the semi-major axis of the orbit. For
$e=0.16$ and $e=0.20$ the average stellar flux is greater by 1.3\% and
2.1\%, respectively, in comparison with a circular orbit. With Gl~581d
located near the outer edge of the HZ, such slightly higher flux may be
precious for the maintenance of habitability. Excursions out of the HZ
may, however, increase the efficiency of irreversible condensation on the
dark side, although these excursions could last only a fraction of the
83-d period.

\subsection{Internal heat flux and tidal dissipation}

 Geothermal heat has a major impact on climate during the early phases
of planetary evolution. Both during and shortly after accretion, it can
trigger a runaway greenhouse and a surface magma ocean provided that the
cooling to space is limited by a dense atmosphere, which would be
blanketing the thermal emission \citep{1988JAtS...45.3081A}. This is due
to the large amount of energy released in a short period of time and to an
efficient convective transport of internal energy to the surface by a yet
liquid mantle. The period during which Earth's surface was significantly
heated ``from underneath'' was restricted to the first tens of Myr.  On
the present Earth, the internal heat flux, dominated by the decay of
radiogenic species and the release of the initial accretion energy, is
about 5000 times lower than the solar energy absorbed by the planet. Thus,
the direct influence of the internal heat on the climate has been
negligible for the past 4.50 Gyrs or so. Similarly, the Gl~581 system is
most likely too old (see Sect.~\ref{sec:erosion}) to invoke geothermal
forcing of the climate. 

The difference in mass and radius between the
Earth and the Gl~581 planets should not affect this
reasoning significantly. The mean flux of accretion energy through the surface scales
with $M^2/R^3$ and remains within 10 times that of the Earth. Assuming the
same abundance of radiogenic species for the Gl~581 planets, the
associated heat flux should scale with $M/R^2$ and remain Earth-like
within a factor of $\sim$2. Only by assuming orders of magnitude more
radiogenic species per unit mass could the geothermal flux be significant
for the global energy balance on the surface.

Interestingly, during the few Myr after the Moon-forming impact, the tidal
interaction with the Moon might have been a major heating source for
Earth's climate \citep{2007SSRv..tmp..127Z}.  Tidal dissipation was
favored by the small Earth-Moon distance and most of the dissipated energy
was taken from the kinetic rotation energy. Because the rotation of Gl~581
planets is presumed to be tidally evolved, tidal dissipation may still
occur but through the damping of their high eccentricity and may increase
the internal heat flux to a level that could affect the climate. It is
thus worth exploring this point further.


Assuming synchronously rotating and eccentric planets, and using formulae
from Peale \citeyearpar{1999ARA&A..37..533P} with a tidal dissipation
factor $Q=100$, we find that $1.4\times 10^{16}$~W and $3.1\times
10^{12}$~W are dissipated respectively in planets Gl~581c and Gl~581d
(these values are obtained with the largest radii given in
Table~\ref{tab:gl581} to analyze the maximum effect). This dissipated
power corresponds to a maximum surface flux of 10~W~m$^{-2}$ for planet
Gl~581c and 0.0015~W~m$^{-2}$ for planet Gl~581d.

On Gl~581c, tidal dissipation can contribute slightly to the energy
balance, with the heat flux being potentially as high as 2\% of the
absorbed stellar energy (about 1\% of the stellar flux at the planet's
distance, averaged over the planetary surface). Besides this direct
warming, this heat flux can increase the volcanic activity and the release
of CO$_2$. At the orbital distance of Gl~581c, additional CO$_2$ can only
further destabilize the climate against water loss or a runaway
greenhouse. Therefore, tidal heating caused by a possibly non-zero
eccentricity speaks against any habitable conditions on this planet.

On planet Gl~581d, tidal dissipation is negligible (0.003\% of the
absorbed stellar energy) and cannot help to directly provide habitable
conditions on its surface despite the low stellar flux. The enhanced heat
flux (compared to that of the Earth) due to both the high planetary mass
and the tidal heating might, however, have indirect effects. By inducing
more volcanism and lasting longer over the planet's history, a higher heat
flux favors the maintenance of the high CO$_2$ level required on this
planet. This may be important when taking into account that the system may
be significantly older than the Earth.

\section{Discussion}
\subsection{Uncertainties in the location of the HZ limits}
Different uncertainties affect the determination of the boundaries of the
HZ. Recently released data on H$_2$O, such as the spectroscopy
and equation of state at high pressure and the temperature and Rayleigh
scattering cross-sections, could induce slight modifications on the
planetary energy budget and should be included in radiative-convective
models. Other data are still lacking, including detailed parameterization
of the CO$_2$-CO$_2$ collision-induced absorption, which is an important
factor for the greenhouse warming close to the outer limit of the HZ.

For H$_2$O- and CO$_2$-rich atmospheres, respectively at the inner and outer edge of the
HZ, the sensitivity of the planetary albedo on the stellar effective
temperature, from which Eqs.~(\ref{eq:fit1}) and (\ref{eq:fit2}) are
derived, was calculated assuming a blackbody spectral distribution of the
stellar flux.
In reality, TiO and H$_2$O absorption makes the spectrum of M-type stars
quite different from a black-body. The calculation should be done for
realistic spectra. For Gl~581, the expected effect would be to slightly
move the HZ boundaries away from the star, because TiO bands mostly block
visible light and shift the emission towards the IR. The albedo
sensitivity to $T_{\rm eff}$ should also be studied for cloudy atmospheres
of planets. Since clouds are a crucial factor in the location of the HZ
boundaries, future simulations will have to include the effect of the
spectral type of the star, in particular for CO$_2$ clouds whose
properties are very sensitive to the wavelength of the incident flux.

An important step forward, especially in predicting the distribution of
clouds, will be the use of 3-D global climate models (GCMs), as was done
to study synchronously rotating planets
\citep{1997joshi,2003AsBio...3..415J} or to address the runaway greenhouse
effect \citep{2002JAtS...59.3223I}.  These models will have to include
realistic microphysical processes for the formation, growth, and
destruction of droplets and icy particles.  To reduce the computing time, 
the treatment of radiative transfer in GCMs has to be greatly
simplified. This is currently the main limitation to the application of
GCMs to exoplanets. At the moment, the detailed line-by-line modeling
required to accurately describe the transfer of radiation in planetary
atmospheres is restricted to 1-D simulations. The development of 3-D GCMs,
including robust treatment of radiative transfer, cloud physics, and
photochemistry, represents the future direction of theoretical studies of
habitability.

\subsection{Geophysical influences} \label{sec:vonbloh}

In Von Bloh et al. (\citeyear{2007arXiv0705.3758V}, hereafter VB07), 
the estimate of the surface temperature is done by solving a system of 
coupled equations, including many geophysical processes affecting the 
CO$_2$ atmospheric level (such as weathering and outgassing rates) and a 
radiation balance equation (Eq. 4 of their paper), written as follows:
\begin{equation}
\frac{L}{4\pi d^2}\left[1 - a(T_{\rm surf},P_{\rm CO_2})\right ] = 4 
I_{\rm R}(T_{\rm surf},P_{CO_2}),
\label{equ:rad}
\end{equation}
where $a$ is the Bond albedo, $I_{\rm R}$ the outgoing thermal emission 
and $d$ the orbital distance. As shown in the present study, 
calculating $a$ and $I_{\rm R}$ requires detailed radiative-convective 
modeling and, besides their dependence on $T_{\rm surf}$ and $P_{\rm 
CO_2}$, $a$ and $I_{\rm R}$ are highly sensitive to the spectral type of 
the star, the cloud properties and cover, and the abundance of other 
radiatively active atmospheric compounds. Solving 
Eq.~(\ref{equ:rad}) numerically requires adopting a simple mathematical form 
for $a(T_{\rm surf},P_{\rm CO_2})$ and $I_{\rm R}(T_{\rm surf},P_{\rm 
CO_2})$, for instance a fit of climate model results. This mathematical 
form and the climate models from which it derives are not given in VB07, 
but most likely they use the same method as the previous works of the 
group \citep[e.g.,][]{2002TellB..54..325F}. This is based on an 
interpolation of radiative-convective results obtained for specific cases 
by Kasting (1993) and Caldeira \& Kasting 
\citeyearpar{1992Natur.360..721C} for a cloud-free atmosphere. This is 
consistent with the HZ inner limit shown in VB07 (figure 3b) 
corresponding (at $t=0$~Gyr) to a radiative-convective calculation with 
$T_{\rm surf}=100^{\circ}$C and no clouds.  There are several points we 
would like to discuss here:
\begin{itemize}
\item[$i)$] A parameterization derived from Kasting et al. (1993) can 
only be applied close to the conditions where the full radiative-convective 
modeling is available, which is close to the inner edge of 
the HZ. The radiative HZ would include Gl~581d only if the effect of CO$_2$ clouds or another greenhouse gas is included. As there are only very few of these calculations at very low stellar irradiation, it would be very useful to know how $a$ and $I_{\rm R}$ are calculated in VB07, in particular 
close to the outer edge of the HZ.
\item[$ii)$] As shown in the present study, clouds have a major but 
uncertain effect on the location of both edges of the HZ. The inner edge, 
for instance, can be located anywhere between 0.05~AU (runaway greenhouse 
with 100\% cover of highly reflective clouds) and 0.11~AU ($100^{\circ}$C, 
no clouds) from the star Gl~581. Therefore, the possible effects of clouds 
have to be compared with the geophysical influences included in VB07.
\item[$iii)$] Although the stellar luminosity is fixed in VB07 (which is a good approximation for M stars), one can see in their Fig.~3  
that the inner boundary of the HZ (fixed at $T_{\rm surf}=100^{\circ}$C) 
moves outward as the planet evolves. Unfortunately, this effect is not explained in the 
paper, so we do not know what geophysical process is involved. 
\end{itemize}

The present study and VB07 present two approaches that should complement 
each other. On one hand, we provide atmospheric models developed to 
compute the radiative transfer and atmospheric structure in a 
self-consistent way. On the other hand, VB07 use a box model including 
geophysical processes relevant for the maintenance of habitable 
conditions. Each of these processes has to be included in the VB07 model 
as a simple mathematical equation (e.g., Eq.~\ref{equ:rad}) to allow the 
numerical calculation of all the variables through a system of coupled 
equations. Although we fully recognize the potential usefulness of this 
approach, we would certainly appreciate an investigation of the 
sensitivity of the VB07 model to all the parameters involved and some 
estimation of the uncertainties affecting the model predictions.

The most noteworthy effect shown in VB07 and previous works from this group is the 
narrowing of the actual HZ compared to the radiative HZ presented for 
instance in our present study. The habitability of a planet can indeed be 
frustrated by the lack of outgassed CO$_2$, as it may have occurred at 
some point in Mars' history. The age at which the decrease of internal 
heat and outgassing makes a planet no longer habitable obviously depends 
on the planetary mass and composition. For a given stellar type, age, and planetary mass, the VB07 model computes the location of the frontier between the inner region 
of the HZ that can actually be searched for habitable planets and the outer 
HZ region that can host only the dead remains of formerly habitable 
planets. However, to define this frontier accurately requires a robust model for each individual geophysical process, as well as strong constraints on the possible variety of planetary compositions; otherwise, we may rule out as ``uninhabitable'' circumstellar regions that could be prime targets in the search for habitable worlds. \\
Two recent publications reach opposite conclusions on 
the onset of plate tectonics on large terrestrial planets. According to 
Valencia et al. \citeyearpar{2007arXiv0710.0699V}, plate tectonics become 
more likely as planets get bigger. O'Neill et al. 
\citeyearpar{2007E&PSL.261...20O} come to the opposite conclusion that the higher gravity of big 
planets tends to prevent the formation of plates, producing a thick and 
unique crust. This debate illustrates the present difficulties in 
extrapolating Earth's geophysical models to exoplanets. Research on these important topics must of course keep going, but it is probably safer at this point to keep the broadest and less specific definition of 
the HZ boundaries from Kasting et al. (1993)\footnote{The precise location of the HZ boundaries based on this definition can of course be revised to account for new molecular data, additional greenhouse agents, 3D calculations, or better cloud models as done, for instance, by Forget and Pierrehumbert (1997), who estimated the possible effect of CO$_2$ clouds in the location of the outer boundary.} and to allow future observations to determine whether a planet found within 
these boundaries actually is habitable.

\section{Conclusion: Are Gl~581c and Gl~581d good targets for Darwin/TPF?}

According to our present knowledge, based on available models of planetary
atmospheres, and assuming that the actual masses of the planets are the
minimum masses inferred from radial velocity measurements, Gl~581c is very
unlikely to be habitable, while Gl~581d could potentially host surface
liquid water, just as early Mars did.

Because of the uncertainties in the precise location of the HZ boundaries,
planets at the edge of what is thought to be the HZ are crucial targets
for future observatories able to characterize their atmosphere. At the
moment, our theory of habitability is only confirmed by the divergent
fates of Venus and the Earth. We will have to confront our models with
actual observations to better understand what makes a planet habitable.
The current diversity of exoplanets (planets around pulsars, hot Jupiters,
hot Neptunes, super-Earths, etc) has already taught us that Nature has a lot
more imagination when building a variety of worlds than we expected from
our former models inspired by the Solar System.

It is obvious that the idealized model of a habitable planet atmosphere,
where the two important constituents are CO$_2$ and H$_2$O, CO$_2$ being
controlled by the carbonate-silicate cycle, is likely to represent only a
fraction of the diversity of terrestrial planets that exist at habitable
distances from their parent star. As an example, planets fully
covered by an ocean may be common, either because they are richer in water
than Earth or because the distribution between surface and mantle water is
different, or perhaps simply because, for a given composition, the mass-to-surface ratio and thus the water-to-surface ratio increases with the
planetary mass, as noted by Lissauer \citeyearpar{1999Lissauer}. Without
emerged continents, it is not at all clear that the carbonate-silicate
cycle could operate. The planets around Gl~581 can fall into this category
since they are significantly more massive than the Earth (especially the
$>$8~$M_{\oplus}$ planet Gl~581d) and also because they may have started
their formation in the outer and more water-rich region of the
protoplanetary disk.

Darwin/TPF-I and TPF-C could eventually reveal what the actual 
properties of the atmosphere of Gl~581c and Gl~581d are. From their thermal 
light curves we could infer if a thick atmosphere is making the climate 
more or less uniform on both the day and night hemispheres of these 
planets, despite a (nearly?) synchronized rotation 
\citep{2004ASPC..321..170S}. Visible and mid-IR water vapor bands could be 
searched in the atmosphere of Gl~581d to confirm its habitability. Mid-IR 
spectra of this planet could also reveal other greenhouse gases at work. 
Spectral observations of Gl~581c could potentially distinguish between a 
Venus-like atmosphere dominated by CO$_2$ or an H$_2$O-rich atmosphere. 
The detection of O$_2$ on this planet would generate a fascinating debate 
about its possible origin: as either a leftover of H$_2$O photolysis and H 
escape or a biological release. There is certainly no doubt that Gl~581c 
and Gl~581d are prime targets for exoplanet characterization missions.

\bibliographystyle{aa}

\end{document}